\documentclass[a4paper]{jpconf} 
\usepackage{graphicx}
\usepackage{listings}
\usepackage{xcolor}

\definecolor{codegreen}{rgb}{0,0.6,0}
\definecolor{codegray}{rgb}{0.5,0.5,0.5}
\definecolor{codepurple}{rgb}{0.58,0,0.82}
\definecolor{backcolour}{rgb}{0.95,0.95,0.92}

\lstdefinestyle{mystyle}{
    backgroundcolor=\color{backcolour},   
    commentstyle=\color{codegreen},
    keywordstyle=\color{magenta},
    numberstyle=\tiny\color{codegray},
    stringstyle=\color{codepurple},
    basicstyle=\ttfamily\footnotesize,
    breakatwhitespace=false,         
    breaklines=true,                 
    captionpos=b,                    
    keepspaces=true,                 
    numbers=left,                    
    numbersep=5pt,                  
    showspaces=false,                
    showstringspaces=false,
    showtabs=false,                  
    tabsize=2
}

\usepackage{lineno}

\modulolinenumbers[1]

\begin{document}
\title{Modern Software Development for JUNO offline software}

\newcommand{\IHEP}{1}
\newcommand{\SDU}{2}
\newcommand{\inst}[1]{$^#1$}

\author{
    Tao Lin\inst{\IHEP}
    (on behalf of the JUNO collaboration)}
\address{
    \inst{\IHEP} Institute of High Energy Physics, Beijing, China. 
}


\ead{lintao@ihep.ac.cn}

\begin{abstract}

The Jiangmen Underground Neutrino Observatory (JUNO), under construction in South China, primarily aims to determine the neutrino mass hierarchy and to precise measure the neutrino oscillation parameters. The data-taking is expected to start in 2024 and the detector plans to run for more than 20 years. The development of the JUNO offline software (JUNOSW) started in 2012, and it is quite challenging to maintain the JUNOSW for such a long time. In the last ten years, tools such as Subversion, Trac, and CMT had been adopted for software development. However, new stringent requirements came out, such as how to reduce the building time for the whole project, how to deploy offline algorithms to an online environment, and how to improve the code quality with code review and continuous integration. To meet the further requirements of software development, modern development tools are evaluated for JUNOSW, such as Git, GitLab, CMake, Docker, and Kubernetes. This contribution will present the software development system based on these modern tools for JUNOSW and the functionalities achieved: CMake macros are developed to simplify the build instructions for users; CMake generator expressions are used to control the build flags for the online and offline environments; a tool named git-junoenv is developed to help users partially checkout and build the software; a script is used to build and deploy the software on the CVMFS server; a Docker image with CVMFS client installed is created for continuous integration; a GitLab agent is set up to manage GitLab runners in Kubernetes with all the configurations in a GitLab repository. 

\end{abstract}

\section{Introduction to JUNO experiment}
The Jiangmen Underground Neutrino Observatory (JUNO) experiment~\cite{Djurcic:2015vqa} has a rich physics program, including the determination of the neutrino mass ordering, precise measurement of neutrino oscillation parameters, detecting neutrinos from reactor, atmosphere, solar, supernova burst, etc~\cite{An:2015jdp,JUNO:2022hxd}. JUNO is under construction in southern China in a underground laboratory, with 700~m overburden (1800 m.w.e.). It is expected to start data-taking in 2024, running for more than 20 years. 

\begin{figure}[h]
    \begin{center}
    \includegraphics[width=0.65\linewidth]{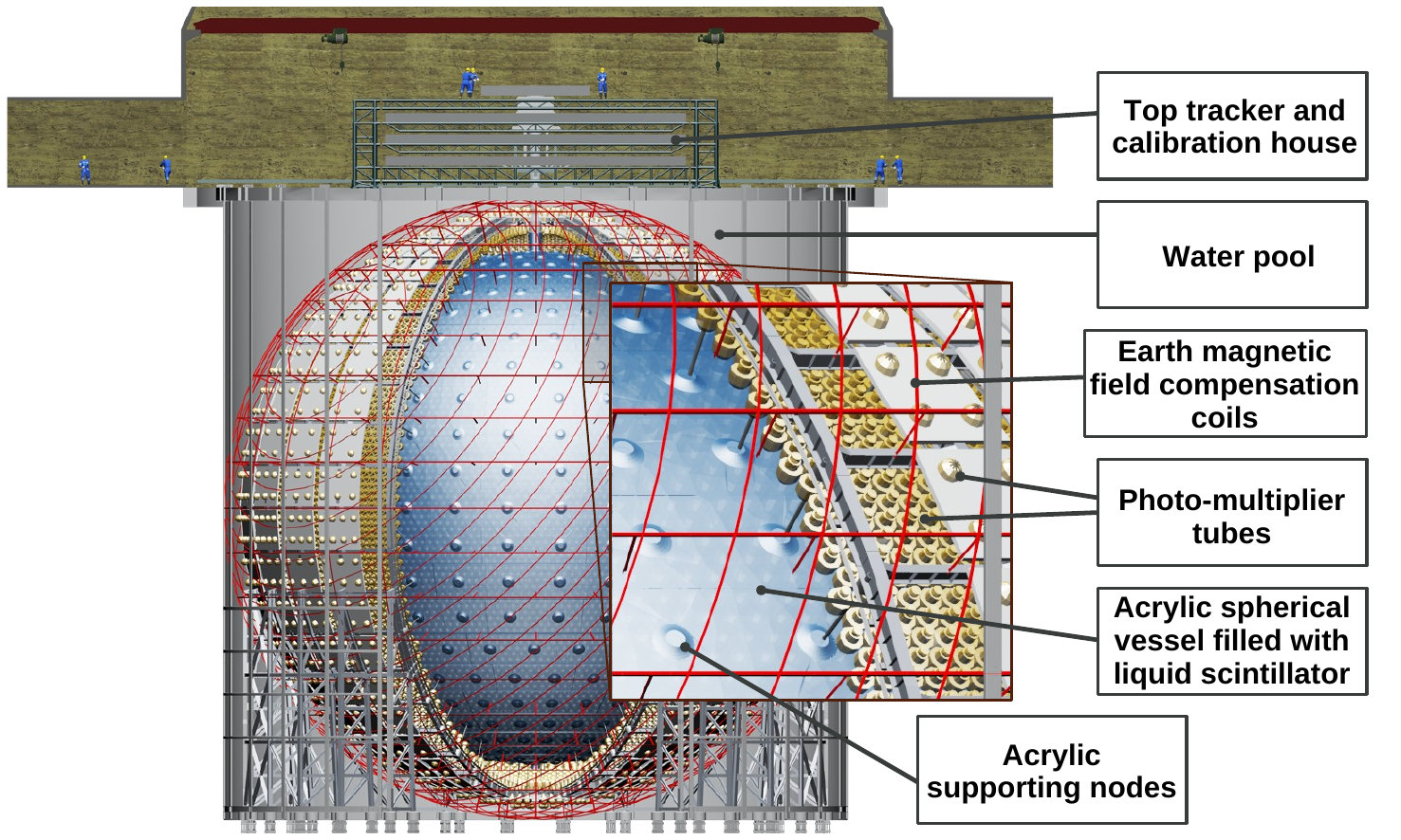}
    \end{center}
    \caption{Schematic view of the JUNO detector.\label{fig:detector}}
\end{figure}

As shown in Figure~\ref{fig:detector}, the JUNO detector consists of a central detector, a water Cherenkov detector, and a top tracker. The innermost part is the central detector with an acrylic spherical vessel filled with 20 kton liquid scintillator (LS), equipped with 17,612 20-inch photomultiplier tubes (LPMT) and 25,600 3-inch photomultiplier tubes (SPMT). The central detector is submerged in a water pool, equipped with 2,400 LPMTs, which is the water Cherenkov detector to detect cosmic ray muons. On the top of the water pool, the top tracker is also used to measure the muons. Further details can be found elsewhere~\cite{An:2015jdp,JUNO:2022hxd}.

\section{Challenges in software development of JUNOSW}
JUNOSW is the offline software for data processing, which is one of the crucial parts of the JUNO experiment~\cite{Lin:2022htc}. It consists of the physics generators, detector simulation, electronics simulation, waveform reconstruction, and event reconstruction. The software is developed based on an underlying framework called SNiPER~\cite{Zou:2015ioy}, whose concept is very similar to the Gaudi framework~\cite{Barrand:2001ny}, which includes event loop, algorithm, service and tool. There is an event loop while the framework is executed. For each event, an algorithm is invoked by the framework to perform a dedicated task. A service provides some common functionalities, which could be invoked by the algorithms. A tool is a piece of code in an algorithm, which improves the code modularity of the algorithm by defining interfaces. All these components are implemented in C++ language and then configured in Python language. 

The development of JUNOSW began in 2012, using Subversion (SVN), Trac and CMT~\cite{Arnault:2000vu} tools. SVN is used for version control and a dedicated web service called Trac is deployed to host the source code. Following the rules of SVN, a SVN repository is created for the JUNOSW, with the three directories: ``trunk'', ``branches'' and ``tags''. The ``trunk'' branch is used for software development. Developers check out this branch and commit their changes back to this branch. The other branches are stored under ``branches'' directory and only used for the preparation of software releases. All the releases are stored under ``tags'' directory. The tool Trac provides a web interface for developers to browse code and submit issues. The tool CMT is used for building the project. The project is organized in packages. Developers could check out dedicated packages with SVN and build with CMT.

The number of packages has increased to more than 200 over the past ten years of development, posing several challenges for both software development and deployment. 
\begin{itemize}
    \item There is a performance issue when building software with CMT. Even though CMT could build a package in parallel, it cannot build different packages at the same time. Building the entire project in a blade server with 28 CPU cores takes about half an hour. This causes the developers to wait for a long time if the project is built from scratch.
    \item Software development lacks code review. When a developer commits the changes to SVN repository, the other developers only receive notifications about the change from a mailing list. Especially when developers add new packages, some binary data could be also committed.  The SVN repository on the server could not be changed, which caused the SVN repository to become large. 
    \item There is a maintenance issue for the continuous integration, which is based on Bitten~\cite{bitten}. Bitten is built on Trac, which only supports Python 2. It consists of a master and several slaves, which need separate deployment. XML-based configuration files need to be set up in the master. When a new commit is pushed to the master, a build task will be created and dispatched to a slave according to the configurations. The slave invokes the commands encoded in the XML when it receives a message from the master. 
\end{itemize}

\section{Adopting modern software development practices for JUNOSW} \label{sec:motivation}
\begin{figure}[h]
    \begin{center}
    \includegraphics[width=0.5\linewidth]{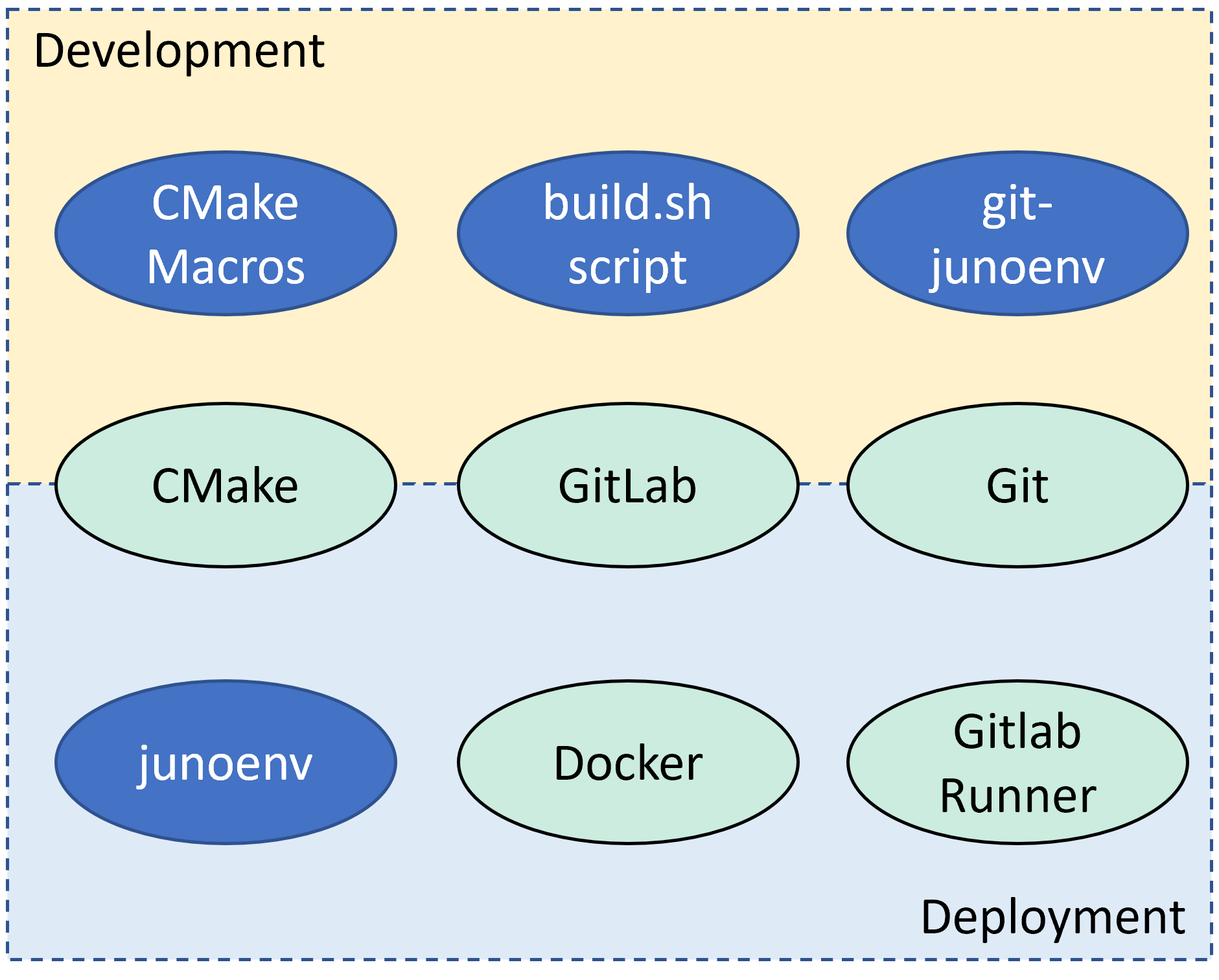}
    \end{center}
    \caption{Overview of tools for software development and deployment in JUNOSW\label{fig:design}. }
\end{figure}

According to the best practices on software development and deployment from HSF (HEP Software Foundation) \cite{Couturier:2017cgq}, modern software tools are adopted by JUNOSW. As shown in figure~\ref{fig:design}, the development and deployment tools are all migrated to modern ones, including CMake, Git/GitLab, Docker and Kubernetes. Meanwhile, some high-level scripts are still developed to help users and developers.

The migration consists of three stages. 
\begin{itemize}
    \item In the first stage, the CMT-based configurations are migrated to the CMake-based. Several CMake macros are developed to help users compile libraries and create setup scripts. With the help of these CMake macros, developers only need to define the name of a package, the dependent targets, and the additional environment variables. 
    \item During the second stage, the repository is migrated from SVN to Git repository. To reduce the size of the repository, the original SVN repository is split into two: one for the source code; and another for the data. Then the latest snapshot of the source code is imported into a new Git repository. The data is put into another Git repository based on Git-LFS (Git Large File Storage). The original histories are also imported into a dedicated Git repository for archival purposes. 
    \item In the third stage, the monolithic project is split into multiple projects. The installation script is moved to a dedicated repository called junoenv. The common packages are moved into a new repository called CommonSW. The CMake macros are modified to handle the dependencies between different projects automatically. In order to support the partial checkout and partial build, a git sub-command named git-junoenv is also developed. 
\end{itemize}

\section{Software development}
\subsection{Migration to CMake}
CMake macros and functions have been developed to put all the common functionalities in the same place. As there are some existing conventions defined in CMT, some of them are used in the CMake macros. According to the instructions in Modern CMake, the following rules have been used in the software development with CMake:
\begin{itemize}
    \item The source code, build directory and installation directory of a project are separated. Even though CMT adopts a similar rule, CMT puts all the build directories under the package directories. When moving to CMake, they are all separated to keep the source code clean. An example is the source code generation for the event data model. When using CMT, the files are generated in the source code directory, which causes the check-in by mistakes sometimes. After moving to CMake, these files are generated under build directories.
    \item A project is organized into packages. A package consists of a header directory for public interfaces, a source directory for the private headers and detailed implementation, a python directory for exporting the library in Python, a share directory for the regularly used scripts, and a test directory for the testing scripts. 
    \item A CMake target is used when building a package. It is used to represent a shared library, a module library or an executable. Its dependencies on the other different packages are described by the other CMake targets. The CMake target properties are used to control how a target is built instead of using global settings. 
    \item CMake generator expression is used to control the flags instead of using the \texttt{if} statement in CMake. This is useful when the same package is built for online and offline environments. In this case, the linking libraries could be different. By using the generator expression, the libraries could be enabled or disabled by checking an option defined in CMake. 
    \item Macros PKG and EDM are developed to build a regular package and a package containing event data model respectively. The macro EDM generates C++ source code and ROOT dictionaries at the CMake configuration stage, and builds a shared library at build stage. The macro PKG creates a shared library by default. If a module library needs to be created, then an option MODULE is needed. If there is no library or executable created, a custom target will be created to install the python and share directories.
    \item Environment variables of packages are collected in the macros PKG and EDM. By adding package names to a global property in the project, all the information of a package could be accessed, including the environment variables. A CMake script is used to create both bash and tcsh scripts before installation. All the environment variables are added at the end of the scripts. 
    \item When a project is installed, a CMake config will be created automatically, including all the targets within the same namespace. Another project needs to use the CMake config file to locate the project and load the exported targets. 
    \item The CMake commands and options are put in the build.sh script. 
\end{itemize}

Below is an example of CMakeLists.txt for the package Geometry:
\begin{lstlisting}[language=bash,basicstyle=\ttfamily\footnotesize]
PKG(Geometry
    DEPENDS
        Identifier
        $<$<NOT:$<BOOL:${BUILD_ONLINE}>>:Parameter>
        Boost::filesystem Boost::system
        Boost::python Python::Python
        ROOT::Geom
    SETENV
        JUNO_GEOMETRY_PATH="$ENV{JUNOTOP}/data/Detector/Geometry"
)
\end{lstlisting}
In this example, the PKG declares the package name. As there are no explicit files to be compiled, all the files under the source code directory will be used. As there is no option MODULE, a shared library will be created. When compiling and building this package, it will depend on several libraries, which are defined after option DEPENDS. As mentioned before, the target names are used. Both Identifier and Parameter are from JUNOSW, while the others are from external libraries. The target Parameter is not used if the software is built for online. 

Below is another example of the build script, which consists of three steps:
\begin{lstlisting}[language=bash,basicstyle=\ttfamily\footnotesize]
function run-build() {
    local installdir=$(install-dir)
    local blddir=$(build-dir)
    check-build-dir
    check-install-dir
    pushd $blddir

    cmake .. $(check-var-enabled graphviz) \
             $(check-var-enabled withoec) \
             $(check-var-enabled online) \
             $(check-var-enabled PerformanceCheck) \
             -DCMAKE_CXX_STANDARD=17 \
             -DCMAKE_BUILD_TYPE=$(cmake-build-type) \
             -DCMAKE_INSTALL_PREFIX=$installdir \
                     || error: "ERROR Found during cmake stage. "
    local njobs=-j$(nproc)
    cmake --build . $njobs || error: "ERROR Found during make stage. "
    cmake --install . || error: "ERROR Found during make install stage. "

    popd
}
\end{lstlisting}

\subsection{Partial checkout and build using git-junoenv}
As Git is already widely used in the HEP community, the migration to Git is not so difficult. After the migration is done, users request to support the partial checkout and build. This is common when using CMT. In order to support partial checkout, git sparse checkout is used. For the partial build, the CMakeLists.txt is set up with customized build targets. 

In order to support the customized build targets, users are allowed to provide their own file, named CMakeLists.user.txt. This file could be edited by users, or controlled by the git-junoenv tool. When using git-junoenv to check out packages partially, these package names will be registered into CMakeLists.user.txt automatically. This could also used to build packages for online environment. Below is an example:
\begin{lstlisting}[language=bash,basicstyle=\ttfamily\footnotesize]
if(BUILD_ONLINE)
  message(STATUS "Using online OEC packages lists")
  include (${CMAKE_SOURCE_DIR}/CMakeLists.online.txt)
elseif (EXISTS "${CMAKE_SOURCE_DIR}/CMakeLists.user.txt")
  message(STATUS "Using user customized package lists")
  find_package(junosw)
  include ("${CMAKE_SOURCE_DIR}/CMakeLists.user.txt")
else()
  message(STATUS "Using default package lists")
  include ("${CMAKE_SOURCE_DIR}/CMakeLists.default.txt")
endif()
\end{lstlisting}

After both partial checkout and build are working, a shell script called git-junoenv is created. By prefixing git in the command name, this command could become a sub-command of git. Below is an example when using it:
\begin{lstlisting}[language=bash,basicstyle=\ttfamily\footnotesize]
$ git junoenv init-project junosw && cd junosw # get the junosw without packages
$ git junoenv list-pkgs                        # list all the available packages
$ git junoenv add-pkg Reconstruction/OMILREC   # add a package
\end{lstlisting}
When users need to develop with JUNOSW, the first step is using init-project to clone the code from the official repository. In order to hide all the packages, both sparse and no-checkout options are used during the git clone. After cloning, the script will check out the CMake-related code and initialize the CMake file of users. Then, users could list all the available packages in the project. The command git ls-files is used to list all the directories containing CMakeLists.txt. Users could use add-pkg to checkout and enable the package.

\section{Software deployment}
\subsection{junoenv: the installation script}
The installation script junoenv is inspired by the ENV project~\cite{env}, which collects all the necessary installation scripts in the same repository. There are more than 50 scripts for building external libraries, and about 30 libraries are deployed in the official release. Modularized bash functions are used to describe the metadata of the external libraries. When installing a package, the script junoenv loads the metadata of the package and drives the installation of it. There are five steps defined during the installation: 
\begin{itemize}
    \item \texttt{get}: download the source code by cURL, wget or git;
    \item \texttt{conf}: configure the package; 
    \item \texttt{make}: build the package;
    \item \texttt{install}: install the package;
    \item \texttt{setup}: create setup scripts for both bash and tcsh.
\end{itemize}
Five corresponding common bash functions are in charge of these steps. If additional configuration is needed, the package can define its own functions to override the default functions. Following is an example:
\begin{lstlisting}[language=bash,basicstyle=\ttfamily\footnotesize]
function juno-ext-libs-cmake-conf- {
    local msg="===== $FUNCNAME: "
    # begin to configure
    echo $msg ./bootstrap --prefix=$(juno-ext-libs-cmake-install-dir)
    ./bootstrap --prefix=$(juno-ext-libs-cmake-install-dir)
}
function juno-ext-libs-cmake-conf {
    juno-ext-libs-PKG-conf cmake
}
\end{lstlisting}
This is used to configure the package CMake. As the CMake does not use the configure script by default, an additional function suffixed with a dash is defined to override the default behavior. So the invoking procedure is: junoenv invokes the conf function of CMake, named \texttt{juno-ext-libs-cmake-conf}; then this function invokes the common function, named \texttt{juno-ext-libs-PKG-conf}; the common function then invokes the overridden function. 

Reproducible is important during the deployment. For a dedicated release of JUNOSW, the versions of all external libraries need to be recorded. In order to track all of them, a shell script is used to collect them. When deploying a release, the corresponding script is loaded. The shell script self is created by invoking the \texttt{vlist} command, which prints all the installed packages and their versions. Below is an example of this script:
\begin{lstlisting}[language=python,basicstyle=\ttfamily\footnotesize]
function juno-ext-libs-git-version- { echo 2.37.3 ; }
function juno-ext-libs-cmake-version- { echo 3.24.1 ; }
function juno-ext-libs-python-version- { echo 3.9.14 ; }
function juno-ext-libs-python-setuptools-version- { echo 58.1.0 ; }
function juno-ext-libs-python-pip-version- { echo 22.2.2 ; }
\end{lstlisting}

All the software is deployed into CVMFS. When deploying the software, the installation prefix could be different from the original one when building software. For most packages, it is not an issue. However, there are still several packages that hardcode the paths, such as physics generators. Inspired by the package manager spack~\cite{10.1145/2807591.2807623}, the paths during building and deployment are set as the same length, and they are replaced using the tool sed when deployed into CVMFS.

\subsection{Docker images for continuous integration}
Continuous integration (CI) is one of the important parts of software development. If building JUNOSW starts from the external libraries, it will take quite a long time. In order to reduce the CI running time, the external libraries should be pre-installed. Two types of Docker images are explored:
\begin{itemize}
    \item Lightweight image with CVMFS clients installed. The image is built based on the CentOS 7. The size of the Docker image after compression is 372.87 MB. This is useful for both CI and developers with good network connections. It is required to run the privileged Docker container so that the CVMFS client can work correctly.
    \item Full image with all the external libraries installed. There are two flavors: one is based on CentOS 7 with a size of 17.29 GB; another is based on Ubuntu 22.04, with a size of 20.37 GB. No special privilege is needed to run the container. 
\end{itemize}

The lightweight image is chosen in the GitLab CI.  The CVMFS client needs to be available before setting up the JUNOSW software environment. Below is the YAML configuration file: 
\begin{lstlisting}[language=bash,basicstyle=\ttfamily\footnotesize]
variables:
  JUNOTOP: /cvmfs/juno.ihep.ac.cn/centos7_amd64_gcc1120/Pre-Release/J23.1.x

default:         # Set the docker image
  image: mirguest/juno-cvmfs

stages:          # List of stages for jobs, and their order of execution
  - build
  - test

build-job-gcc:   # This job runs in the build stage, which runs first.
  stage: build
  script:
    - sudo mount -t cvmfs juno.ihep.ac.cn /cvmfs/juno.ihep.ac.cn
    - source $JUNOTOP/setup.sh
    - ./build.sh
\end{lstlisting}

The benefit is that the Docker image could be reused when upgrading the external libraries. In the above example, only the \texttt{JUNOTOP} needs to be updated in the YAML file. 

\subsection{GitLab runners in Kubernetes cluster}
GitLab runners are in charge of the execution of the CI jobs. They need to be deployed and associated with the GitLab projects. As there are multiple projects, GitLab group runners are set up in a self-hosted Kubernetes cluster. GitLab agents are used to connect Gitlab and Kubernetes. Then the Gitlab runners are managed by the Gitlab agents. All the configurations are managed in GitLab repositories. 

The GitLab agent is set up as below. 
\begin{lstlisting}[language=bash,basicstyle=\ttfamily\footnotesize]
gitops:
  manifest_projects:
  - id: JUNO/offline/gitlab-agent
    default_namespace: junooffline
    paths:
    - glob: 'manifests/*.{yaml,json}'
    - glob: '/**/*.{yaml,json}'

ci_access:
  groups:
  - id: JUNO/offline
\end{lstlisting}

Then GitLab runners are installed with the cluster management project, which is a Git repository. The configuration of GitLab runner is enabled first, then the runners will be set up automatically. 

\begin{lstlisting}[language=bash,basicstyle=\ttfamily\footnotesize]
repositories:
- name: gitlab
  url: https://charts.gitlab.io

releases:
- name: runner
  namespace: gitlab-managed-apps
  chart: gitlab/gitlab-runner
  version: 0.44.0
  installed: true
  values:
    - values.yaml.gotmpl
\end{lstlisting}

\section{Conclusions}
The tools used for JUNO software development have been migrated from CMT and SVN to CMake, Git/GitLab, Docker and Kubernetes. Some additional scripts have been also developed to help the users. In late 2022, all the migration had been done. More than 160 members are available in the JUNO GitLab. In the past nine months, more than 300 Merge Requests have been merged into the official JUNOSW repository, and more than 2,400 pipelines have been executed with a success ratio of 92.04\%. 

\ack
This work is supported by National Natural Science Foundation of China (12375195, 12025502, 11805223), the Strategic Priority Research Program of the Chinese Academy of Sciences (Grant No. XDA10010900), and Youth Innovation Promotion Association, CAS.

\section*{References}
\bibliography{iopart-num}

\end{document}